\documentstyle[11pt,newpasp,twoside,epsf]{article}
\def\edcomment#1{\iffalse\marginpar{\raggedright\sl#1\/}\else\relax\fi}
\reversemarginpar

\begin{document}
\title{Present Status of Diffusive Shock Acceleration}
 \author{Hyesung Kang}
\affil{Department of Earth Sciences, Pusan National University, Pusan,
609-735, Korea}

\begin{abstract}
Diffusive shock acceleration (DSA) is now widely accepted as the model
to explain the production of cosmic rays (CRs) in a wide range 
of astrophysical environments.  Despite initial successes of the 
theory in explaining the energetics and the spectrum of CRs 
accelerated by supernova remnants, there still remain some 
unresolved issues such as particle injection out
of the thermal plasma at shocks, CR diffusion due to the self-generated
MHD waves and yet-to-be-detected gamma-ray emission due to the ionic
CRs.  Recent technical advancements to resolve these issues are reviewed.
\end{abstract}

\section{Introduction}

The origin of cosmic rays had been by and large an unsolved
mystery until 1980s, but during last two decades we have put together
many pieces of this important astrophysical puzzle. We now believe
most of galactic cosmic rays, at least up to $10^{14}$ eV of the
particle energy, are accelerated by the supernova blast waves
within our Galaxy (Drury 1983; Blandford \& Eichler 1987; Jones et al. 1998).
As reviewed in Jones (2001), the DSA theory is beautifully constructed
and explains successfully many observational aspects of the galactic CRs.

In many respects DSA theory is a quantitative science that makes 
detailed predictions consistent with observations:  
1) roughly speaking, order of 10 \% of shock energy can be transferred to CRs
in case of supernova blast waves, 
2) the accelerated particle distribution has a power-law spectrum with 
some modifications due to non-linear feed back,
3) CR composition can be also explained to the lowest order by the acceleration 
of average interstellar medium convolved with propagation and spallation 
effects. 
Unfortunately, pion-decay gamma-ray emission 
from specific sources has not been positively identified,
although we can fit the observed radiation emitted by CR electrons with 
reasonable assumptions on model parameters. 
Of course more realistic, free-of-free parameter calculations will 
help us answer to the question if DSA theory really works in real 
physical shocks. 

This paper is organized as follows.
We will first review the numerical simulation studies of DSA theory 
and the observational evidences of particle acceleration in various 
astrophysical shocks.
Next we will describe some unresolved issues of DSA theory, in particular,
injection process and diffusion models.  
Then we will introduce some recent efforts to improve the numerical 
techniques designed to handle these problems.
Finally we will briefly mention what lies ahead in terms of both 
theoretical and observational researches.

\section{Application of DSA Theory}
Since we now believe most of galactic cosmic rays
are accelerated by the supernova blast waves, DSA at supernova remnants (SNR) 
has been received the most attention. 
There are several physical reasons why we believe SNRs are the acceleration 
sites of galactic cosmic rays:

\begin{enumerate}
\item{} 
First of all, we know for sure from the direct measurements at Earth's 
bow shock (Ellison et al. 1990) and interplanetary shocks (Baring et al. 1997)
that the particles do get accelerated at the shock. 
Also plasma simulations (Quest 1988) show that
ions can be scattered back and forth across the shock by self--generated
waves, and that these scattered
ions can provide a seed population of cosmic rays.

\item{} 
Secondly, SNRs are the most energetic phenomena occurring in our 
Galaxy and only candidates to explain the energy requirement for the constant 
CR energy density. According to numerical studies, order of 10 \% of SN 
explosion energy can be transferred into cosmic rays, although the exact 
fraction depends on the detailed model parameters (Jones \& Kang 1992;
Berezhko et al 1994).  
Considering that one SN goes off about every 30 years in our Galaxy, 
the energy injection rate into CRs from SNRs is about $10^{41}$ erg/s.  
This can replenish the energy of CRs escaping from our Galaxy 
(Blandford \& Eichler 1987).

\item{} 
According to the standard acceleration model where we assume a spherically 
symmetric shock propagating into the uniform ISM and a mean magnetic field 
parallel to the shock normal and the Bohm diffusion model, the maximum 
energy of the particles that can be accelerated by a typical SNR 
is about $10^{14}$ eV for protons (Lagage \& Cesarsky 1983).  
In oblique shocks, the particles may be accelerated to even higher energy 
due to cross-field diffusion and drift acceleration (Jokipii 1987).  
Also if one includes the nonlinear effect, that is, much higher compression 
and greater velocity jump across the shock, the maximum energy is increased.
(Berezhko 1996)

\item{} 
Also according to numerical studies of Berezhko et al.(1994), 
the accelerated particle spectrum at the SNR is a power-law of index of -2.1 
at the source. 
After considering the propagation through ISM and escape process, the 
expected spectrum at the Earth steepens to $E^{-2.7}$ which is close to 
what we observe.  
\end{enumerate}

We note here Biermann (1993) suggested 
that SNRs exploding inside wind-blown bubbles can accelerate the heavy ions 
up to the Ankle energy with energy independent diffusion due to macroscopic 
turbulences. 
Stanev, Biermann \& Gaisser (1993) predicted the observed particle 
spectrum which is proportional to $E^{-2.75}$ for protons dominating 
the spectrum below the Knee, while to $E^{-3.07}$ for heavy ions 
dominating above the Knee. In this model, the energy dependence of the 
escape time scale is derived from Kolmogorov type magnetic turbulences.

\subsection{Review of Numerical Studies}
The concept behind DSA of charged
particles trapped between convergent flows across a shock, is quite
simple. However, the full DSA problem
is actually extremely complex, because the nonlinear
interactions between energetic particles, resonantly scattering waves and
the underlying plasma can become dominant effects.
Important consequences of nonlinear interactions include such things as
generation and damping of the scattering wave field, injection of suprathermal
particles into the CR population, as well as
heating and compression of the plasma flow due to the CR pressure.
Because of these complex nonlinear physics, numerical simulations 
have been primary tools to study the details of the acceleration process 
and dynamical feedback of the CRs to the underlying plasma
(Kang \& Jones 1991; Berezhko et al. 1994; Berezhko \& V\"olk 2000).

Among various numerical methods, I will mention only the following three 
techniques (for additional background see also Jones 2001).

\subsubsection{Monte Carlo Method}
In Monte Carlo simulations, one follows the scattering of individual
particles, based on an assumed scattering law, by the underlying flow 
around a one-dimensional shock which is assumed to be in a steady-state. 
Using Monte Carlo Simulations, Ellison et al. (1990) 
calculated the particle spectra accelerated in quasi-parallel portion of 
Earth's bow shock and successfully compared them with observational data.
They showed that the agreement between simulation results and observed 
data was quite impressive. But the highest energy accelerated by the
shock goes only up to 100 keV due to small size of the Earth's bow-shock.
They also showed the results of Monte Carlo simulations were consistent
with those of hybrid plasma simulations.
Baring et al. (1997) also did the same kind of comparison with the observed 
data in oblique interplanetary shocks and also came up with excellent 
agreements.

\subsubsection{Two-Fluid Model}
CR acceleration at SNRs are simulated first by two-fluid method in which CR
energy density is solved instead of the distribution function  
(Drury \& Falle 1986; Kang \& Jones 1990; Dorfi 1990; Jones \& Kang 1992).
The main conclusion was that order of 10 \% of SN explosion energy can be 
transferred to CRs with reasonable assumptions on the closure parameters 
and injection rate. 
But it was realized at the same time that the final outcomes are 
sensitively dependent on the closure parameters, the adiabatic index of CRs,
$\gamma_c$ and the momentum averaged diffusion coefficient, $<\kappa>$,
which are free parameters of the model we must assume a priori.
However, as long as we adopt injection rates and the closure parameters
inferred from the diffusion-advection equation calculations,
the acceleration efficiency and the shock structure calculated
with the two-fluid method are in good agreement with those computed with
the diffusion-advection method (Kang \& Jones 1995).

\subsubsection{Kinetic simulations}
The next generation numerical method was kinetic simulations in which the 
diffusion-convection equation for the distribution function $f(p)$ is solved
(Kang \& Jones 1991; Berezhko et al. 1994; Berezhko \& V\"olk 
2000).
Within this method, we can eliminate one of the closure parameters, that is, 
the adiabatic index of CRs, $\gamma_c$.  
Also instead of a momentum averaged diffusion coefficient, a more realistic, 
momentum-dependent diffusion model, $\kappa(p)$ can be adopted.  
But the injection rate still remains as a free parameter.
Berezhko and collaborators 
have extensively studied various aspects of 
DSA model for SNRs in a series of papers.
They showed that, for SNR in uniform hot ISM, 
the energy transferred to CR component is about 20 \% of total SN energy.
In early stage, the accelerated particle spectrum is the test-particle
like power-law spectrum, but nonlinear modification affects the spectrum
significantly later on. Yet the SNR shock structures never become smooth, 
owing to the geometrical factors in an expanding spherical shock 
(Berezhko et al. 1994). 
Although the particle spectrum at shock becomes a concave curve when
a significant precursor (``foot''-like structure upstream to shock) develops, 
the overall spectrum of CR protons integrated over the entire volume is 
a power-law with the index of 2.1, $E^{-2.1}$ up to $10^14$ eV.
In Berezhko \& Volk (2000), SN type II remnants exploding into a wind-blow 
bubble were also considered with the similar results as in a uniform
background. 
The main conclusion of these simulations was that DSA can be very 
efficient for strong shocks if we assume Bohm diffusion and the 
injection rate of $10^{-4}$ to $10^{-3}$.
So DSA simulations by Berezhko's group have presented quantitative 
predictions that can explain many aspects of the galactic CRs. 
Thus they place SNRs as the prime candidate for the acceleration 
sites of the galactic CRs.

\subsection{Observational Evidences for DSA}

There are many direct and indirect evidences for the particle 
acceleration at various astrophysical shocks
(Blandford \& Eichler 1987).

\subsubsection{Interplanetary Shocks}
We mentioned in the previous section direct measurements of the particle 
acceleration process at Earth's bow shock and interplanetary shocks. 
Those measurements are reproduced reasonably well by various numerical 
techniques such as Monte Carlo simulations, hybrid plasma simulations, 
and kinetic simulations.

\subsubsection{Supernova Blast Waves}
Moving on to larger scale and to stronger shocks, let's look at SNRs.
Most successful observations have been made for CR electrons.  
Many SNRs are observed by radio synchrotron radiation due to relativistic 
electrons gyrating around a magnetic field.  
In some remnants this synchrotron radiation extends to X-rays. 
For example, Koyama et al. (1995) detected X-ray synchrotron emission 
in SN1006 with ASCA X-ray telescope. 
SN1006 emits synchrotron X-ray radiation at two bright rims 
whose X-ray energy spectrum is a power-law, a typical signature of 
synchrotron emission. 
The maximum electron energy emitting this radiation was 
estimated to be 100 TeV, which is consistent with what the 
standard model predicts. 
Relativistic electrons can be detected also in gamma rays by nonthermal 
Bremsstrahlung and by inverse Compton scattering of cosmic microwave 
background radiation.  The TeV gamma rays from SN1006 detected by 
Cangaroo experiment is believed to be due to 
the inverse Compton scattered CBR from 40 TeV electrons (Tanimori et al 1998).
We have not been so lucky, however, in detecting CR protons in SNRs.  
Relativistic protons collide with the ISM and emit gamma rays via pion decay  
(i.e. $p + p \rightarrow \pi^0 \rightarrow \gamma~ ray$).
According to theoretical estimates, the detection of pion decay gamma rays 
from nearby SNRs may be difficult with current detectors but not impossible
(Drury et al. 1994, Berezhko \& Volk 2000).  
So far most of gamma ray observations gave only upper limits, 
but no positive detections yet.
This failure for proton $\gamma$ ray detection calls for further
improvements on theoretical modeling and numerical calculations
as well as experimental sensitivity.
We note also there is an intrinsic difficulty in distinguishing pion decay
gamma rays from electron IC gamma rays, since often IC gamma rays
dominate in some remnants as in SN1006.

\subsubsection{Radio Galaxies}
Moving onto even larger scale than SNRs, we have radio galaxies 
whose radio emission is thought be synchrotron radiation due to CR electrons. 
The strongest emission comes from so-called ``hot spots'' that are working
surfaces where the jet flow strongly interacts with the IGM.
Since electron synchrotron loss time is very short compared to the travel 
time along the jet, the particles at the lobes need to be accelerated locally
(Begelman et al. 1984).
As we heard in earlier talks (Biermann 2001), powerful radio galaxies could 
be the origin of UHECRs above GZK energy. 

\subsubsection{Intracluster Medium}
Recently clusters of galaxies have played important roles in cosmology, 
because their properties and distribution can provide important clues 
to large scale structure formation (Bahcall 1999).  
Accordingly, clusters have been 
actively observed not only in X-ray and optical bands, but also in radio 
and in EUV. 
There are more than 25 clusters that have diffuse radio halos (Giovannini
et al. 1999). 
The radio emission is once again due to CR electrons.
The possible origins of non-thermal electrons include:
1) re-acceleration of CR electrons previously ejected by radio galaxies 
inside a cluster.
2) secondary electrons generated by interactions of CR protons with 
background Intracluster medium.
3) freshly injected electrons by merger shocks and cosmic structure 
shocks associated with the bulk flows due to large scale structure formation.

The existence of CR electrons are also suggested through observations of 
IC scattering of CMBR in hard X-ray (Fusco-Femiano et al.
2000; Feretti et al. 2000) and EUV (Lieu et al. 1996).
One of most important outcomes of having both synchrotron and IC scattering 
observations is that we can estimate the magnetic field strength without 
resorting to the energy equipartition argument.
Although direct observations reveal only the presence of CR electrons
so far, one generally assumes that protons are present with at least
comparable numbers and greater energy content (Lieu et al. 1999).
Thus it may indeed be reasonable to expect a substantial energy
in CR protons in the ICM can be dynamically important 
in formation and evolution of clusters. 
Also recent observation by Clarke, Kronberg, \& B\"ohringer (2000) suggested
that the mean magnetic field inside ICM may be as high as 5 microgauss,
which is certainly dynamically significant.
Although these new observations need to be looked at more carefully,
we believe that cosmology community is beginning to appreciate possible
roles of CRs in the evolution of the Universe (Miniati et al. 2000)

\section{Unresolved Issues}

In the kinetic version of diffusive shock acceleration theory, we solve 
the diffusion convection (DC) equation for CR distribution function along with 
the usual gas dynamics equations including the contribution of CR pressure.
The diffusion--convection equation which describes the time evolution of
the particle distribution $f(p,x,t)$ (e.g.~Skilling 1975)
takes the form:
\begin{equation}
\frac{df}{dt}~=~\frac{1}{3}
      (\frac{\partial u}{\partial x})
       p\,\frac{\partial f}{\partial p}+ \frac{\partial }{\partial x}
        \left(\kappa(x,p)\frac{\partial}{\partial x}f\right) + Q \,.
\end{equation}
The diffusion coefficient $\kappa(p)$ and the injection rate $Q$ from thermal
particles to CRs are the primary free parameters in this model.
Thus most of uncertainties in the DSA theory lie in the magnetic field
configuration and MHD wave spectrum which determines the particle injection
process at the shocks and the diffusion model.
According to plasma simulations of quasi-parallel shocks, the streaming 
motion of the CR particles against the background fluid can induce Alfven 
waves that scatter the particles (Quest 1988).  
Then the suprathermal particles are 
scattered and injected into CRs efficiently at the parallel shocks.  
On the other hand, in perpendicular shocks, both self-generation of 
waves and particle injection may become inefficient.
So more quantitative calculations are necessary to determine how 
the particles are injected and accelerated in perpendicular or oblique shocks.
 
Especially for electrons whose gyro-radius is much smaller than ionic 
gyro-radius some additional process is needed to bridge the gap between
the thermal electron population and the relativistic region.
Recently energy transfer from waves amplified by ions reflected off the
shock has been suggested as a pre-acceleration (electron injection)
mechanism (Levinson 1996; McClements et al. 1997).
 
\subsection{The Injection Process}
In early numerical simulations (Kang \& Jones 1991, Berezhko et al. 1994), 
the particle injection is realized by putting a fixed fraction, $\eta$, of 
incoming particles at an injection momentum, $p_{inj}= \lambda m_p c_{s,2}$, 
where $\lambda \sim 2$ and $c_{s,2}$ is the sound speed of postshock gas.
Thus injection process is controlled by two free parameters, injection 
rate and the ratio of the injection momentum to thermal momentum.

This model is refined further in the so-called ``thermal leakage'' type 
injection model where small fraction of suprathermal particles in the 
high energy tail of the Maxwellian distribution diffuses upstream and 
then gets injected into CRs. The injection momentum is defined as
$p_{inj} = c_1\,2\sqrt{m_p k_{\rm B} T_2}$,
where $k_{\rm B}$ is the Boltzmann's constant and
$T_2$ is the postshock gas temperature.
In Kang \& Jones (1995), the particles above the injection momentum 
are allowed to diffuse according to D-C equation solver, 
while the distribution just 
below the injection momentum is assumed to be Maxwellian defined by 
the local temperature.  So the injected particle number flux is no 
longer a free parameter, but rather it is connected with thermal 
distribution in an explicit way as $\eta = 1.6 c_1^3 \exp(-2c_1^2)$. 
So if one fixes the parameter $c_1$, the injection rate $\eta$ is determined 
by this relation.  
Kinetic simulations with this thermal leakage injection 
have produced the results consistent with both observed data and the 
results of Monte Carlo simulations for quasi-parallel Earth's bow shock 
and oblique interplanetary shocks (Kang \& Jones 1995; Kang \& Jones 1996).

\subsection{Diffusion Model}
The most important physical quantity in DSA theory is perhaps the spatial 
diffusion coefficient which quantifies the complex interactions between 
the particles and waves in the magnetic field.
Both analytic studies and plasma simulations have shown that Alfven waves 
are excited by CRs streaming upstream of parallel shocks
(Skilling 1975; Bell 1978; Lucek \& Bell 2000).
So even if there are not enough irregularities to begin with, especially 
upstream to the shock, scattering waves can be self-generated by CRs 
diffusing upstream. 
Then these waves are amplified by the shock and advected to downstream.
So a common practice is to assume these self-generated waves provide 
scattering strong enough so that the particles are scattered roughly
in one gyro-radius. 
So the Bohm diffusion is regarded as a reasonable assumption for a 
quasi-parallel shock.
But it remains uncertain if spherically expanding SNR shocks can 
generate strong scattering waves for the highest energy particles 
upstream to the shock.
 
The mean magnetic field configuration 
can vary from quasi-parallel to quasi-perpendicular in both ISM SNRs 
and wind-bubble SNRs.
It is uncertain if strong waves would be excited at all in 
quasi-perpendicular shocks.
So if one assumes the Bohm diffusion coefficient for SNR shocks, 
it only represents an upper limit of particle acceleration.
We need to understand in more details the complex physics involved 
in the particle scattering and diffusion, especially in perpendicular shocks.

\section{Recent Developments in Numerical Techniques}

\subsection{Self-Consistent Injection Scheme}
Recently Malkov (1998) has presented a self-consistent, analytic, 
nonlinear calculations 
for ion injection based on the interactions of the suprathermal particles 
with self-generated MHD waves in strong shocks.
He calculated a transparency function which expresses the fraction of 
suprathermal particles that are able to leak upstream through the waves.
By adopting this analytic solution, Gieseler et al. (2001)  have developed a 
numerical treatment of the injection model at a strong quasi-parallel shock.
In this scheme, the transparency function is approximated by 
\begin{equation}
\tau_{esc}(v,u_2)~=~H\left[ \tilde{v}-(1+\epsilon) \right]
        \left(1-\frac{u_2}{v}\right)^{-1}\,
        \left(1-\frac{1}{\tilde{v}}\right)\nonumber\\
 ~ ~ \exp\left\{-\left[\tilde{v}-(1+\epsilon)\right]^{-2}\right\}\,,
\end{equation}
which depends on the postshock flow speed, $u_2$,  particle speed,
$v$, and the wave amplitude parameter, $\epsilon$.
The condition ${\partial \tau_{esc}(p)}/{\partial p} \not= 0 $ in fact
defines the ``injection pool'' where the thermal leakage takes place.
The transparency function is used as a filter, so the probability for 
leakage is zero below the injection pool, while the leakage probability 
is unity above the injection pool. 
Thus the particle momentum in the injection pool 
just above the Maxwellian tail are injected with a certain probability.
This injection scheme eliminated the last remaining free-parameter 
in the Kang \& Jones (1995) model; that is, an injection momentum parameter $c_1$, 
and so it can provide a self-consistent injection scheme without 
free parameters except the parameter $\epsilon$ which 
is well constrained from plasma simulations.
According to the simulations by Gieseler et al (2000),
it turns out that the injection process is {\it self-regulated} in such a way
that the injection rate reaches and stays at a nearly stable value after
quick initial adjustment, but well before the CR shock reaches a steady 
state.
The effective value of injection parameter $c_1$ is about 2.3 and the 
injection rate is about $10^{-3}$ for both Mach 30 and Mach 2 shocks.
Long term evolutions for shocks of a wide range of Mach numbers should 
be simulated with a more cost-effective code described below.

\subsection{Shock-Tracking AMR code}
The main difficulty in numerical techniques to solve nonlinear diffusive 
shock acceleration is the fact the diffusion-convection equation 
includes an extremely wide range of length scales that need to be resolved.  
For a Bohm type diffusion where scattering length is proportional to the 
particle momentum, the diffusion length increases linearly with the momentum,
that is, $l_{diff} \propto p$.
So if we consider the acceleration from supra-thermal particles ($p_{th}/m_pc 
\sim 10^{-3}$) to the Knee energy ($p_{max}/m_pc \sim 10^6$), for example, 
the ratio of largest length scale to the 
smallest length scale to be resolved is $l_{max}/l_{min}\sim 10^9$. 
 
So far there is only one code by Berezhko et al. (1994) that can handle such 
a strong momentum dependent diffusion model in time-dependent simulations.  
They introduced a ``change of variables technique'' in which the radial 
coordinate is transformed into a new variable scaled with particle 
diffusion length as $x(p)= \exp(-(r-R_s)/l_{diff}(p))$ where
$R_s$ is the shock radius. 
This allowed them to solve the coupled system of 
gasdynamic equations and the CR transport equation even when the diffusion 
coefficient has a strong momentum dependence.  
As we mentioned earlier, their code is different from conventional 
numerical codes in several ways.
Gasdynamic equations and the CR transport equation are solved separately 
both downstream and upstream side of the gas subshock.
Then the gasdynamic solutions at both side of the subshock are used to 
solve the Riemann problem, which determines how the subshock evolves.
Also an iteration scheme is applied to match the downstream and upstream 
solutions for the CR diffusion-convection equation at the subshock.
In any case this has enabled them to explore several important issues 
regarding the particle acceleration at supernova remnants more fully than 
was possible before.
Berezhko and Ellison (1999) have compared the results from this code and 
Monte Carlo simulations with reasonable agreements, 
but it still needs to be tested more thoroughly against other conventional 
numerical methods.

So we have developed a new CR shock code that can perform kinetic 
simulations with a strong momentum-dependent diffusion (Kang et al. 2001).
Here the basic features of this code will be introduced.
Total transition of a CR modified shock consists of a subshock and 
a precursor just upstream to the subshock. 
In order to follow accurately the evolution of a CR modified shock, it is 
necessary to resolve the precursor structure (Kang \& Jones 1991). 
At the same time, in order to calculate the particle injection it is also 
necessary to solve correctly the diffusion of the low energy suprathermal 
particles at the shock (Kang \& Jones 1995).
To satisfy these two requirements, we need to refine only the small region 
around the shock with higher resolution grids.
Hence our new code combines the so-called Adaptive Mesh Refinement technique 
with the shock-tracking scheme.

\begin{figure}
\plotone{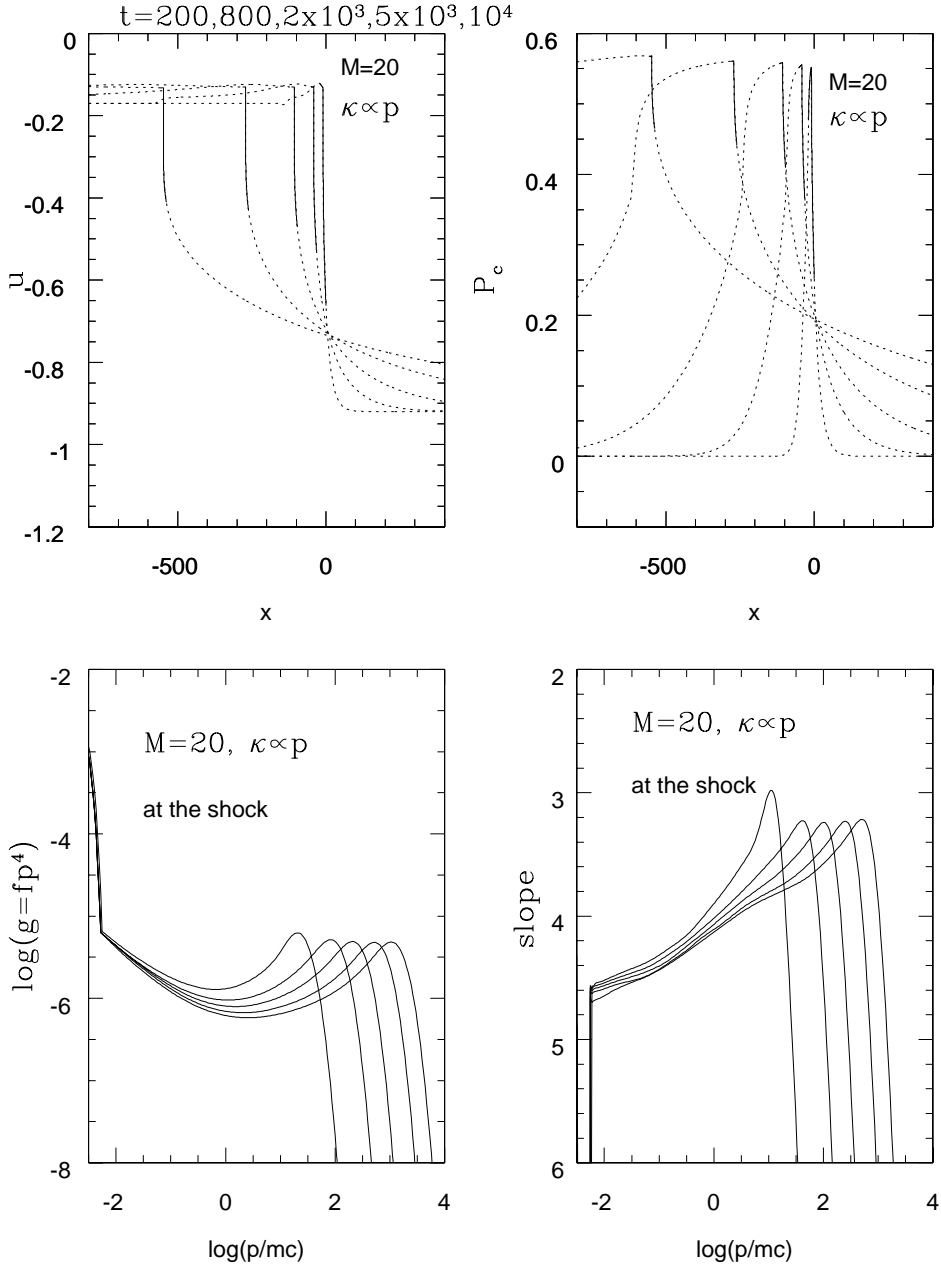}
\caption{
Time evolution of the flow velocity and the CR pressure for
the $M=20$ shock with $\kappa_l \propto p$ until $t=10^4$,
simulated by the shock-tracking/AMR code.
Also the CR distribution function $g=f(p)p^4$ at the shock and 
its power slope $q=-\partial \ln{f} / \partial \ln{p}$.
Five levels of refined grids are used in addition to the base grid.
The shock is slowly drifting to the left in the simulation frame.
}
\end{figure}

In the shock tracking method of Le Veque \& Shyue (1995), the underlying base grid 
has uniform cells. An additional cell boundary is introduced at the 
location of the shock, subdividing a uniform cell into two sub-cells.
In the next time step, this cell boundary is moved to a new location 
according to the Riemann solutions and the hydrodynamic waves are propagated 
onto the new set of grid zones.
Since the new grid is chosen so that the shock wave coincides exactly with 
the cell boundary, the shock remains as an exact discontinuity without 
smearing.
This shock tracking scheme is implemented to the adaptive mesh refinement 
method of Berger \& Le Veque (1997).
Aided by the information about the exact shock location, we can refine 
the region around the shock with multiple levels of grids.
A fixed number of cells around the shock are identified as 
the ``refinement region'' on the base grid. 
A typical cell number for the refinement regions is about 200.
The 1st level refined grid is generated by placing twice more cells 
within the refinement region, so each cell is refined by a factor of two.
Then the half of the cells around the shock on the 1st level grid are 
chosen to be refined further to the 2nd level grid, making the length 
of the refinement region a half of that in the 1st level grid.  
The same refinement procedure is applied to higher level grids.

Fig. 1 shows the time evolution of a Mach 20 shock with a Bohm 
type diffusion coefficient.  Five levels of refined grids in addition 
to the base grid are used. 
Here dashed lines show the structures in the base grid, while the solid 
lines show the structures in the 1st refined grid. 
The flow velocity and CR pressure are shown in order to highlight the 
development of the precursor.
The bottom two panels show the distribution function and its slope as a 
function of particle momentum.  The distribution function shows typical 
concave curves reflecting non-linear modification.  
In the simulated shock, the compression ratios across the subshock and
across the total transition are 3.1 and 11, respectively, so
$f(p)$ is $\propto p^{-4.5}$ at low energy momenta but flattens
to $ f(p) \propto p^{-3.3}$ at high energy momenta just below $p_{\rm max}$.
This demonstrates that nonlinear feedbacks between the precursor dynamics
and the CR injection and acceleration should be treated accurately in
numerical simulations of CR shocks.

Using this code, we can save both computational time and memory over what 
would be required to solve the problem using more traditional methods 
on a single fine grid.
In typical simulations where 10\% of the base grid is refined with $l_{\rm
max}$ levels, for example,
the computing time increases by factors of $(2^{l_{\rm max}})^{0.7}$
compared with the case of no refinement ($l_{\rm max}=0$).
It should be compared with the time increases by factors of
$ (2^{l_{\rm max}})^2$ for the simulations of an uniform grid spacing that
matches the cell size at the $l_{\rm max}-th$ refined level grid.
With the shock-tracking/AMR code we can use the Bohm type diffusion 
and follow the injection of suprathermal particles at the same time 
in CR shock simulations.
After the self-consistent injection model of Gieseler et al. (2000) is
implemented into this code, it will provide a powerful numerical tool
to study the CR injection and acceleration at astrophysical shocks. 

\section{Future Studies}

Here we will summarize with a brief discussion on what needs to be 
improved in future studies. 
We need to refine further the numerical calculations of the particle 
acceleration at SNRs in order to make more realistic predictions on 
the proton spectrum and pion decay gamma ray radiation flux.

\begin{itemize}
\item{}
Our shock-tracking AMR code equipped with the self-consistent 
injection model described in the previous section will enable 
us to do more realistic time-dependent 
simulations of CR shocks. It will also provide a way to confirm various 
calculations done by Berezhko' s group with a different method.
\item{}
A self-consistent model for diffusion can be adopted, if the wave 
energy density is included in the hydro/CR code as the third component.
\item{}
Pre-acceleration of electrons needs to be studied further, perhaps with 
3D plasma simulations.
\item{}
In terms of observation, improved data for spectra and composition 
around the Knee will help us constrain better DSA models for the origin 
of galactic CRs. 
\item{}
Also better understanding of the properties of SNRs, that is,
magnetic field configuration, surrounding density distribution, composition, 
and properties of different SN type will help us make more quantitative 
predictions from SNR simulations.
\item{}
Finally, positive detection of pion decay gamma-rays from SNRs will give us
a clear proof for acceleration of CR protons at SNRs.
\end {itemize}

\acknowledgements
This work was supported by Korea Research Foundation Grant
(KRF-2000-015-DP04448).


\begin{references}
Bahcall, N. A. 1999, in {\it Formation of Structure in the Universe}, 
eds: A. Dekel  \&  J. Ostriker, Cambridge:
Cambridge University Press, pp. 135-171.

\reference Baring, M. G. et al. 1995, Adv. Space Res. 15, 397

\reference Begelman, M. C., Blandford, R. D. \& Rees, M. J. 1984, 
Rev. Mod. Phys., 56, 255 
)
\reference Bell, A. R. 1978, \mnras, 182, 147 

\reference Berezhko E.G. 1996, Astropart. Phys., 5, 367 

\reference Berezhko E.G., Yelshin V.K., \& Ksenofontov L.T. 1994,  
Astropart. Phys., 2, 215 

\reference Berezhko E.G., \& Ellison, D. C. 1999, \apj, 526, 385 

\reference Berezhko E.G., \& V\"olk H.J. 2000, \aap, 357, 283 

\reference Berger, J. S., \& LeVeque, R. J. 1997, SIAM J. Numer. Anal. 

\reference Biermann, P. B. 1993, \aap, 271, 649 

\reference Blandford R.~D., \& Eichler D. 1987, Physics Reports, 154, 1 

\reference Clarke, T. E., Kronberg, P. P.  and B\"ohringer, H. 2000, 
astro-ph/0011281

\reference Dorfi, E.A., 1990, \aap, 234, 419

\reference Drury L.~O'C. 1983, Rep. Prog. Phys., 46, 973 

\reference Drury, L.~O'C., \& Falle, S.~A.~E.~G. 1986, \mnras, 223, 353

\reference Drury, L. O'C., Aharonian, F., \& Volk, H. J. 1994, \aap, 287, 959 

\reference Ellison, D.~C., M\"obius, E., \& Paschmann, G. 1990, \apj, 352, 376

\reference Feretti, L. et al. 2000, in proceedings of IAP 2000 Conference 
"Constructing the Universe with Clusters of Galaxies" (Paris), astro-ph/0009346

\reference Fusco-Femiano, R. et al. 2000, ApJL, 534, L7

\reference Gieseler U.D.J., Jones T.W., \& Kang H. 2001, \aap, in press, 
astro-ph/0011058

\reference Giovannini, G., Tordi, M., Feretti, L. 1999, New Astronomy, 4, 141

\reference Jokipii, J.R. 1987, \apj, 313, 842 

\reference Jokipii, J.R., \& Morfill, G. 1987, \apj, 312, 170 

\reference Jones, T. W. 2001, in these proceedings 

\reference Jones, T. W., \& Kang, H. 1992, \apj, 396, 575 

\reference Jones, T. W., et al. 1998, \pasp, 110, 125

\reference Kang H., \& Jones T.W. 1990, \apj, 363, 499 

\reference Kang H., \& Jones T.W. 1991, \mnras, 249, 439 

\reference Kang H., \& Jones T.W. 1995, \apj, 447, 944 

\reference Kang H., \& Jones T.W. 1997, \apj, 476, 875 

\reference Kang H., Jones T.W., LeVeque, R., \& Shyue, K. M. 2001, \apj, 
 March 20 issue, astro-ph/0011538

\reference Koyama, K. et al. 1995, Nature, 378, 255 

\reference Lagage, P.O., \& Cesarsky, C.J. 1983, \aap, 118, 223 

\reference Lee, M.A. 1982,  J. Geophys. Res., 87, 5063 

\reference Lieu, R. et~al. 1996, \apjl, 458,  5 
 
\reference Lieu, R., Ip, W.-H., Axford, W. I. \& Bonamente, M. 1999, \apjl,
510, 25 

\reference LeVeque, R. J., \& Shyue, K. M. 1995, SIAM J. Scien. Comput. 16, 348

\reference Levinson A. 1996, \mnras, 278, 1018 
 
\reference Lucek \& Bell, 2000, \mnras, 314, 65

\reference Malkov M.A. 1998, Phys. Rev. E,  58, 4911

\reference McClements K.G., Dendy R.O., Bingham R., Kirk J.G., Drury L. O'C.
1997, \mnras, 291, 241 

\reference Miniati, F., Ryu, D., Kang, H., Jones, T.~W., Cen, R., \& 
Ostriker, J.~P. 2000, \apj,  542, 608


\reference Quest K.B. 1988, J. Geophys. Res.,  93, 9649

\reference Skilling J. 1975, \mnras, 172, 557 

\reference Stanev, T., Bierman, P. L., \& Gaisser, T. K. 1993, \aap, 274, 902 

\reference Tanimori, T., et al. 1998, \apjl, 497, L25 

\end{references}
\end{document}